\documentclass[12pt,draftcls,journal,onecolumn]{IEEEtran}
\ifCLASSINFOpdf
  \usepackage[pdftex]{graphicx}
  \graphicspath{{../pdf/}{../jpeg/}}
  \DeclareGraphicsExtensions{.pdf,.jpeg,.png}
\else
  \usepackage[dvips]{graphicx}
  \graphicspath{{../eps/}}
  \DeclareGraphicsExtensions{.eps}
\fi
\usepackage{color}
\usepackage{epsfig}
\usepackage{booktabs}
\usepackage{epstopdf}
\usepackage{amssymb}
\usepackage{mathtools}
\usepackage{algorithmic}
\usepackage{array}
\usepackage{mdwmath}
\usepackage{mdwtab}
\usepackage{bm}
\usepackage{mathrsfs}
\usepackage{epsfig}
\usepackage{subfigure}
\usepackage{algorithmic}
\usepackage{amsmath}
\usepackage{mathrsfs}
\usepackage{cite}
\usepackage{url}

\DeclareGraphicsExtensions{pst,eps}
\graphicspath{{fig/}}

\begin{document}
\title{Transfer Learning and Meta Learning Based Fast Downlink Beamforming Adaptation}
\author{Yi~Yuan,
       ~Gan~Zheng,~\IEEEmembership{Senior Member,~IEEE,}
       Kai-Kit Wong,~\IEEEmembership{Fellow,~IEEE,}\\
        Bj\"{o}rn Ottersten,~\IEEEmembership{Fellow,~IEEE,}
        Zhi-Quan Luo,~\IEEEmembership{Fellow,~IEEE}
\thanks{Y. Yuan and G. Zheng are with the Wolfson School of Mechanical, Electrical and Manufacturing Engineering, Loughborough University, Loughborough, LE11 3TU, UK (E-mail: \{y.yuan, g.zheng\}@lboro.ac.uk).}
 \thanks{K.-K. Wong is with the Department of Electronic and Electrical Engineering, University College London, London, WC1E 6BT, UK (Email: kai-kit.wong@ucl.ac.uk).}
 \thanks{B. Ottersten is with Interdisciplinary Centre for Security, Reliability and Trust (SnT), University of Luxembourg, L-1359 Luxembourg (Email: bjorn.ottersten@uni.lu).}
 \thanks{Z.-Q. Luo is with Shenzhen Research Institute of Big Data, and the Chinese University of Hong Kong, Shenzhen, China (Email: luozq@cuhk.edu.cn).}
 }
\markboth{\it To appear in IEEE Transactions On Wireless Communications}{}
\maketitle
\vspace{-2cm}
\begin{abstract}
This paper studies fast adaptive beamforming optimization for the signal-to-interference-plus-noise ratio balancing problem in a multiuser multiple-input single-output downlink system. Existing deep learning based approaches to predict beamforming rely on the assumption that the training and testing channels follow the same distribution which may not hold in practice. As a result, a trained model may lead to performance deterioration when the testing network environment changes. To deal with this task mismatch issue, we propose two offline adaptive algorithms based on deep transfer learning and meta-learning, which are able to achieve fast adaptation with the limited new labelled data when the testing wireless environment changes. Furthermore, we propose an online algorithm to enhance the adaptation capability of the offline meta algorithm in realistic non-stationary environments. Simulation results demonstrate that the proposed adaptive algorithms achieve much better performance than the direct deep learning algorithm without adaptation in new environments. The meta-learning algorithm outperforms the deep transfer learning algorithm and achieves near optimal performance. In addition, compared to the offline meta-learning algorithm, the proposed online meta-learning algorithm shows superior adaption performance in changing environments.
\end{abstract}
\vspace{-0.5cm}
\section{Introduction}
Multi-antenna techniques have been widely used to improve the spectral efficiency of modern wireless communications systems  due to their ability to exploit spatial characteristics of the propagation channel \cite{lim2013recent,boccardi2014five}. Beamforming is recognized as one of the most promising multi-antenna techniques since it can efficiently improve the antenna diversity gain and mitigate multiuser interference. In the last two decades, beamforming optimization has been well studied for some specifical problems, such as signal-to-interference-plus-noise
ratio (SINR) balancing problem \cite{schubert2004solution,bengtsson1999optimal}, power minimization problem \cite{rashid1998transmit,wiesel2005linear} and sum rate maximization problem \cite{bengtsson1999optimal,shi2011iteratively,zheng2014joint,yuan2019joint}. Most beamforming design problems are solved using either tailor-made iterative algorithms or general iterative algorithms using convex optimization tools. However, iterative algorithms may have slow convergence. This fact causes severe computational latency and makes the optimized beamforming solutions outdated. Hence, existing beamforming techniques have difficulty meeting the demands for real-time applications in the fifth generation (5G) systems. Although heuristic methods such as zero-forcing (ZF) beamforming are faster to implement, they often show far from optimal system performance. Hence, designing efficient solutions that balance computational complexity and performance  has attracted much attention.

Recently, deep learning (DL) has been recognized as an efficient technique to solve difficult design problems in wireless communications due to its ability of modeling highly non-linear functions at considerably lower complexity \cite{zhang2019deep,wang2017deep,zappone2019wireless}. Accordingly, DL techniques have been widely used in many applications of wireless networks to address specific physical layer issues, such as channel estimation and decoding \cite{he2018deep,ye2017power,liang2018iterative}, hybrid precoding \cite{huang2019deep,elbir2019hybrid,elbir2019deep} and resource allocation \cite{sun2017learning,lee2018deep,liang2019towards}. The successful application of the DL techniques on the problems of resource allocation \cite{sun2017learning,lee2018deep,liang2019towards} is based on the learning to optimize framework, which aims to learn a simple mapping through the deep neural network (DNN) instead of optimizing the complex mathematic problems. Motivated by the above successful applications of DL techniques, it is possible to address the tradeoff issue between complexity and performance in the beamforming design. This is the result of the mapping from the input channel state to output beamforming that is obtained by training the neural networks in an offline manner. The beamforming solution can be directly predicted using the trained network in real time. The advantage of the learning to optimize framework is to transfer the complex real-time optimization procedures to offline training showing great potential to solve the beamforming design problems in multi-antenna systems \cite{alkhateeb2018deep,shi2018learning,huang2018unsupervised,xia2019deep,zhang2020deep}. In \cite{alkhateeb2018deep}, a DL model was proposed to predict the beamforming matrix with a restricted codebook and a finite solution space which cause performance loss. In order to improve the performance, the works in \cite{shi2018learning,huang2018unsupervised} directly predicted the beamforming matrix using the trained network. However, the direct prediction method may cause high training complexity and low learning accuracy of the neural networks since the number of variables to predict increases significantly as the number of transmit antennas and users increases. To overcome this drawback, the authors in \cite{xia2019deep} exploited the problem structure and proposed a model-based DL framework to optimize the beamforming matrix. The proposed model-based framework includes two parts: the DL part used to learn the optimal mapping from the channel to the uplink power allocation as key features with much reduced dimension than the original beamforming matrix, and the signal processing part used to recover beamforming from the predicted uplink power allocation. By utilizing the specific problem structure, a DL enabled approach was proposed to optimize beamforming of the SINR balancing problem under per-antenna power constraints \cite{zhang2020deep}. The proposed DL algorithms in \cite{alkhateeb2018deep,shi2018learning,huang2018unsupervised,xia2019deep,zhang2020deep} are based on a common assumption that the training and testing channel data are drawn from the same distribution in a fixed stationary environment. However, this assumption may be violated in real-world systems due to the dynamic nature of wireless networks. As a result, existing DL based optimization algorithms may cause a task mismatch issue when the network environment changes. A straightforward way is to re-train the model from scratch using newly collected data for each new network environment. However, this method results in huge overhead of data collection and training time. Hence, overcoming the task mismatch issue in deep learning to optimize beamforming becomes a major challenge in dynamic communications environments.

Transfer learning is a promising technique to deal with the task mismatch issue experienced in the practical wireless communication systems due to its ability to transfer the useful prior knowledge to a new scenario \cite{pan2009survey}. The basic idea of transfer learning is to extract the key features of the source domain and   refine the pre-trained model in the target domain. The efficiency of the transfer learning technique on solving the task mismatch issue has been investigated in the resource allocation of wireless communications \cite{zappone2019intelligence,shen2018transfer}. Another efficient way to deal with the task mismatch issue is meta-learning, which aims to improve the learning ability by leveraging the different but related training and testing data  \cite{thrun1998learning}.  Most existing meta-learning algorithms are problem-specific. In order to eliminate the model architecture restriction on the applications of meta-learning, the authors in \cite{finn2017model} proposed the model-agnostic meta-learning (MAML) algorithm. The MAML algorithm aims to learn a parameter initialization of the model for fast adaptation by alternating between inner-task procedure and cross-task procedure. Specifically, the task-specific parameters are updated by performing the gradient descent on the loss function of the corresponding task in the inner-task procedure, and the global network parameter is updated by performing the gradient descent on the sum of the loss function of the associated tasks in the cross-task procedure. Based on the advantages of the MAML algorithm on solving mismatch issues, it has been used to deal with the channel estimation problems in wireless communication systems \cite{park2019learning,park2019meta,yang2019deep}. For instance, MAML-based meta-learning algorithm was proposed to solve the decoding problem over fading channels \cite{park2019learning}, to estimate the end-to-end channel with insufficient pilots \cite{park2019meta}, and to predict channel state information (CSI) of frequency division duplexing systems \cite{yang2019deep}. The simulation results in \cite{park2019learning,park2019meta,yang2019deep} indicate that meta-learning is able to achieve better adaptation performance compared to the joint training method since the joint training method uses the overall available data in the source domain and target domain to train the model without the adaptation process.

Although transfer learning and meta learning techniques have been used to solve the  channel estimation and decoding problems \cite{park2019learning,park2019meta,yang2019deep}, they are still in the early stage for wireless communications applications. Different from channel estimation and decoding problems, beamforming design is a well-known challenging problem and there is no known solution to the optimal adaptive beamforming in a dynamic wireless environment. Hence, it is important to design adaptive beamforming algorithms to solve the mismatch issue. In addition, directly applying adaptive learning techniques to solve the high dimensional beamforming solution will cause high training complexity and inaccurate results. In order to improve the accuracy and reduce the complexity of neural network training, we choose the uplink power allocation vector as the low dimensional feature to predict. To be specific, we propose the offline and online fast adaptive algorithms using transfer learning and meta learning techniques to solve the mismatch issue of beamforming design in dynamic wireless environments. Our main contributions are summarized as follows:
\begin{itemize}
\item We propose an offline adaptive learning algorithm based on deep transfer learning (DTL) by combining DL techniques and transfer learning to achieve the adaption to a new environment. This algorithm first trains a model in the source domain which includes   channel data different from those in the testing environment. It then refines the pre-trained model by fixing common feature layers and re-training the fully connected   layer in the target domain which includes few labelled data from the testing environment.  
\item We proposed an offline adaptive learning algorithm based on meta-learning by utilizing the idea of MAML. This algorithm includes two parts: the meta-training part and the fine-tuning part. The meta-training aims to optimize the global parameter initialization via alternating between the inner-task procedure and the cross-task procedure using data in the source domain. The fine-tuning part refines the initialized parameter using the global parameter and limited data in the new environment. The advantages of the proposed meta-learning include fast adaptation and near optimal performance.
\item We propose an online adaptation algorithm  to further improve the adaptation capability of the  offline meta-learning algorithm in the real-world non-stationary communications scenarios where the environment constantly changes such that new labelled data only arrive sequentially. This algorithm is designed based on meta-learning and the `follow the leader (FTL)' method. The FTL method is used to deal with the sequential data in real-time systems  and the meta-learning method is used for fast adaptation.
\item Extensive simulations are carried out to evaluate the adaption capability of the proposed algorithms in realistic communications scenarios using WINNER II and 3GPP channel models. The results verify the adaption performance of the proposed offline algorithms and indicate that the offline meta-learning algorithm can achieve near optimal performance by avoiding the huge data collection and training time in new communications scenarios. In addition, the proposed online algorithm can significantly improve the adaption of the offline meta-learning algorithm in non-stationary scenarios.
\end{itemize}

The remainder of this paper is organized as follows. Section \ref{system_model} introduces the system model and the beamforming neural network (BNN) learning framework. In section \ref{offline_adaptation}, the offline DTL algorithm and meta-learning algorithm are proposed. Section \ref{online_adaptation} develops the online meta-learning based adaptation algorithm. Simulation results and conclusions are presented in Section \ref{simu} and Section \ref{conc}, respectively.

{\em Notions:} The boldface lower case letters and capital letters are used to represent column vectors and matrices, respectively. The notation $\mathbf{a}^H$ and $\|\mathbf{a}\|_2$ denote the Hermitian conjugate transpose and the $l_2$-norm of a vector $\mathbf{a}$, respectively. The operator $\mathcal{CN}(0, \mathbf{\Theta})$ represents a complex Gaussian vector with zero-mean and covariance matrix $\mathbf{\Theta}$. $\mathbf{I}_M$ denotes an identity matrix of size $M\times M$. Finally, $\leftarrow$ denotes the assignment operation.
\section{System Model}\label{system_model}
A multi-input   single-output (MISO) downlink transmission system is considered, in which $K$ single antenna users are served by a base station (BS) with $M$ antennas. The received signal at user $k$ can be expressed as
\begin{align}
\mathbf{y}_k = \mathbf{h}_k^H\mathbf{w}_k\mathbf{s}_k+n_k,~\forall k
\end{align}where $\mathbf{h}_k\in\mathbb{C}^{M\times1}$ denotes the channel coefficient between the BS and user $k$, $\mathbf{w}_k$ and $\mathbf{s}_k\sim\mathcal{CN}(0,1)$ denote the transmit beamforming and the information signal for user $k$, respectively, and the additive Gaussian white noise (AWGN) is given by $n_k\sim\mathcal{CN}(0,\sigma_k^2)$. Consequently, the signal-to-interference-plus-noise ratio (SINR) balancing problem can be formulated as:
\begin{align}\label{SINRproblem}
\max_{\mathbf{w}_k,k=1,\ldots,K}~\min_{1\leq k\leq K}~\frac{|\mathbf{h}_k^H\mathbf{w}_k|^2}{\sum_{j\neq k}^K|\mathbf{h}_k^H\mathbf{w}_j|^2+\sigma_k^2}, ~~~~\mathrm{s.t.}~~\sum_{k=1}^K\|\mathbf{w}_k\|^2\leq P,
\end{align}where $P$ is power budget. Many existing algorithms can be used to generate the optimal solution of problem \eqref{SINRproblem}. Although the existing DL-based algorithms can solve the issue of outdated beamforming caused by the conventional optimization algorithms, they will cause the task mismatch issue when the network environment changes. Hence, we focus on the design of fast adaptive learning algorithms to overcome the task mismatch issue in beamforming design in dynamic network environments.

\begin{figure}[ht]
\centering
\includegraphics[width=3.8in]{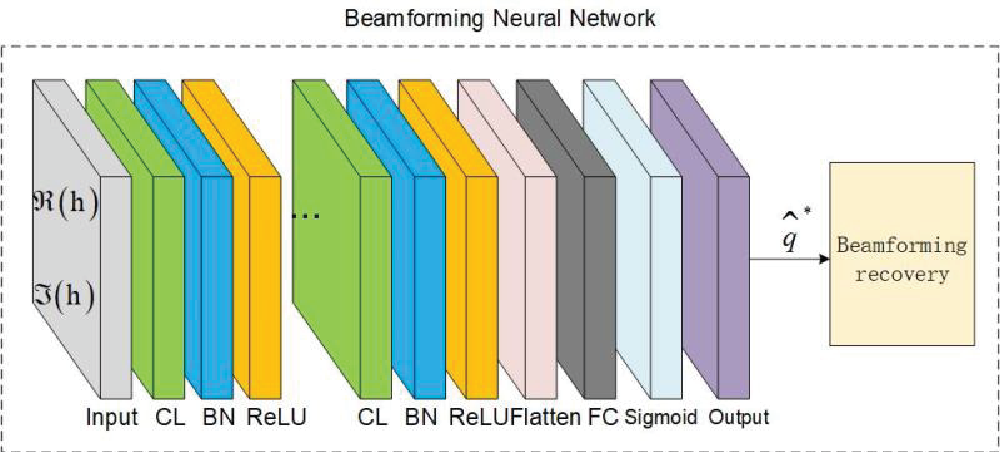}
\caption{The DL-based BNN for uplink power prediction and beamforming recovery \cite{xia2019deep}.}
\label{bnn}
\end{figure}
Directly predicting beamforming causes high training complexity and inaccurate results due to the high dimensional beamforming matrix, so instead we predict the low dimensional uplink power allocation vector. According to the uplink-downlink duality in \cite{schubert2004solution} and \cite{xia2019deep}, using the uplink power allocation vector to replace beamforming as the output of the neural network is possible because the same SINR region of the uplink and downlink problems can be achieved. Based on uplink-downlink duality and normalized beamforming $\mathbf{w}_k=\tilde{\mathbf{w}}_k\sqrt{p_k}$, the downlink problem \eqref{SINRproblem} can be converted into the following uplink problem
\begin{align}\label{uplproblem}
\max_{\mathbf{q}}~\min_{1\leq k\leq K}~\frac{q_k|\mathbf{h}_k^H\tilde{\mathbf{w}}_k|^2}{\sum_{j\neq k}^Kq_j|\mathbf{h}_j^H\tilde{\mathbf{w}}_k|^2+\sigma_k^2}, ~~~~\mathrm{s.t.}~~\|\mathbf{q}\|_1\leq P,~\|\tilde{\mathbf{w}}_k\|_2=1,\forall k,
\end{align}where $\mathbf{q}=[q_1,\ldots,q_K]^T$ and $q_k$ is the uplink power allocation for user $k$, $\tilde{\mathbf{w}}_k$ and $p_k$ are the normalized beamforming and downlink power allocation of user $k$, respectively. Sufficient labelled data can be generated by solving problem \eqref{uplproblem}. The model-based BNN approach proposed in \cite{xia2019deep} is chosen for our algorithms design since it can efficiently extract features and can recover the high dimensional beamforming matrix from the low dimensional feature vector. The BNN framework shown in Fig. \ref{bnn} includes two modules: the neural network module and the beamforming recovery module. First, we introduce how to recover beamforming matrices by using the recovery module. With the predicted uplink power allocation vector, the normalized beamforming vector can be obtained as $\tilde{\mathbf{w}}_k=\frac{(\sigma_k^2\mathbf{I}+\sum_{k=1}^Kq_k\mathbf{h}_k\mathbf{h}_k^H)^{-1}\mathbf{h}_k}
{\|(\sigma_k^2\mathbf{I}+\sum_{k=1}^Kq_k\mathbf{h}_k\mathbf{h}_k^H)^{-1}\mathbf{h}_k\|_2},\forall k$. Then, the optimal downlink power allocation vector $\mathbf{p}=[p_1,\ldots,p_K]^T$ can be obtained by finding the first $K$ components of the eigenvector of the following matrix
\begin{align}
\mathbf{\Upsilon}(\tilde{\mathbf{W}}, P)=\left[
\begin{matrix}
\mathbf{D}\mathbf{U}&\mathbf{D}\mathbf{\sigma}\\
\frac{1}{P}\mathbf{1}^T\mathbf{D}\mathbf{U}&\frac{1}{P}\mathbf{1}^T\mathbf{D}\mathbf{\sigma},
\end{matrix}\right],
\end{align}where $\mathbf{1}=[1,1,\ldots,1]^T$, $\mathbf{D}=\mathrm{diag}\{1/|\tilde{\mathbf{w}}_1^H\mathbf{h}_1|^2,\ldots,1/|\tilde{\mathbf{w}}_K^H\mathbf{h}_K|^2\}$, $\mathbf{\sigma}=[\sigma_1^2,\sigma_2^2,\ldots,\sigma_K^2]^T$, and $[\mathbf{U}]_{kk^{'}}=|\tilde{\mathbf{w}}_{k^{'}}^H\mathbf{h}_k|^2$, if $k^{'}=k$, otherwise $[\mathbf{U}]_{kk^{'}}=0$. Finally, the downlink beamforming matrix $\mathbf{W}=[\mathbf{w}_1,\ldots,\mathbf{w}_k]$ is derived as $\mathbf{W}=\tilde{\mathbf{W}}\sqrt{\mathbf{P}}$, where $\tilde{\mathbf{W}}=[\tilde{\mathbf{w}}_1,\ldots,\tilde{\mathbf{w}}_K]$ and $\mathbf{P}=\mathrm{diag}(\mathbf{p})$. Second, we briefly describe the neural network framework used in the paper according to the neural network module of BNN. The convolutional neural network (CNN) architecture is chosen as the base of the learning framework in this paper due to its ability of extracting features and reducing learned parameters. Specifically, the CNN framework includes the input layer, convolutional layer (CL), batch normalization (BN) layer, activation (AC) layer, and fully connected layer (FC). Channel realization is split into two real value inputs. One is the in-phase component $\mathfrak{R}(\mathbf{h})$ of channel realization and the other one is the quadrature component $\mathfrak{I}(\mathbf{h})$ of channel realization.

As a regression problem is considered in this paper, we use supervised learning and the standard mean squared error (MSE) as the loss function to calculate the loss of the neural network. The loss function is defined as follows:
\begin{align}\label{loss}
\mathrm{Loss_\mathbb{D}}(\mathbf{\theta})=\frac{1}{N}\sum_{i=1}^N\|\hat{\mathbf{q}}^{(i)}(\mathbf{\theta})-\mathbf{q}^{(i)}\|^2_2,
\end{align}where $\mathbb{D}=\{(\mathbf{h}^{(i)},\mathbf{q}^{(i)})\}_{i=1}^N$ is the training dataset, $\mathbf{q}^{(i)}$ and $\hat{\mathbf{q}}^{(i)}(\mathbf{\theta})$ denote the optimal uplink power allocation vector generated by solving \eqref{uplproblem} and the predicted uplink power allocation vector of the neural network for the $i$-th sample in each batch, respectively, $\mathbf{\theta}$ is the network parameter  and $N$ is the batch size. In the following sections, we will design our fast adaptive learning algorithms.

\section{Offline Learning Algorithms}\label{offline_adaptation}
In this section, we design two offline adaptive learning methods to optimize beamforming: 1) DTL algorithm and 2) meta-learning algorithm. These two algorithms aim to achieve fast adaptation in the new test wireless environment with limited channel data whose distribution is different from that in the training environment. In the following subsections, we describe the details of these two algorithms.
\subsection{Joint Training}
In this subsection, we introduce the joint training method, which is considered as a benchmark for evaluating the adaptation ability of the proposed offline algorithms. The joint training method aims to learn a single model on a joint dataset. Hence, the objective of the joint training method can be expressed by the following optimization problem
\begin{align}
\mathbf{\varphi}=\arg\min_{\mathbf{\varphi}}\mathrm{Loss}_{\mathbb{D}_{joint}}(\mathbf{\varphi}),
\end{align}where $\mathbb{D}_{joint}$ denotes the joint training dataset, which is generated by merging the training data and adaptation data, and $\mathbf{\varphi}$ is the parameter vector of the single model. The parameter vector $\mathbf{\varphi}$ can be iteratively updated based on the following gradient-based learning rule
\begin{align}\label{jointupdate}
\mathbf{\varphi}\leftarrow\mathbf{\varphi}-\alpha\nabla_{\mathbf{\varphi}}\mathrm{Loss}_{\mathbb{D}_{joint}}(\mathbf{\varphi}),
\end{align}where $\alpha$ is the learning rate.
\subsection{Deep Transfer Learning}
Transfer learning has been recognized as an efficient method for model prediction as it not only reduces the dependence on a large amount of labelled data but also avoids training the model from scratch. Different transfer learning methods and corresponding applications have been introduced in \cite{pan2009survey}.
Since the same SINR balancing optimization problem are used to guide beamforming design over different wireless environments, it indicates that some common features
inherent in the optimization problem can be extracted and transferred. Therefore, a DTL algorithm via fine-tuning, which combines the DL and transfer learning techniques, is proposed to generalize the beamforming prediction in different channel distributions.
\subsubsection{Definition  of Datasets}
The fundamental idea of transfer learning is to train the neural network in a given source domain and then adapt the model to a target domain. To apply DTL, we first define datasets for the network model. We use the algorithm in \cite[Table 1]{schubert2004solution} to generate $N_{Tr}$ sample pairs for each user to compose the training dataset $\mathbb{D}_{\mathrm{Tr}}(\cdot)$, which will be used to create the pre-trained model. Then, we use the same process to generate the adaptation dataset $\mathbb{D}_{\mathrm{Ad}}(\cdot)$ with $N_{Ad}$ sample pairs using the test channel fading distribution different from that used in generating $\mathbb{D}_{\mathrm{Tr}}(\cdot)$. We assume that any sample pair in the testing dataset $\mathbb{D}_{\mathrm{Te}}(\cdot)$ does not appear in the adaptation dataset $\mathbb{D}_{\mathrm{Te}}(\cdot)$.
\subsubsection{Transfer Learning}
The proposed DTL includes two stages: 1) building the pre-trained neural network model in the source domain 2) refining the pre-trained model in the target domain. In the first stage, we minimize the loss function $\mathrm{Loss_{\mathbb{D}_{Tr}}}(\mathbf{\theta})$ on the training dataset $\{\mathbb{D}_{\mathrm{Tr}}(k)\}_{k=1}^{N_{Tr}}$, which includes sufficient sample pairs, to optimize the network model. The network parameter can be updated by using the following equation
\begin{align}\label{tranparaupdate}
\mathbf{\theta}\leftarrow\mathbf{\theta}-\alpha\nabla_{\mathbf{\theta}}\mathrm{Loss_{\mathbb{D}_{Tr}}}(\mathbf{\theta}),
\end{align}where $\alpha$ is the network learning rate, $\nabla_{\mathbf{\theta}}\mathrm{Loss_{\mathbb{D}_{Tr}}}(\mathbf{\theta})$ is the gradient of the loss function over $\mathbf{\theta}$. Alternatively, the network parameter $\mathbf{\theta}$ also can be updated by using the existing adaptive moment estimation (ADAM) algorithm \cite{kingma2014adam}.

Next, we move to the fine-tuning stage when the pre-training stage is finished. Fine-tuning aims to refine all or partial parameters of the pre-trained neural network on the target task. To fast adapt to the new environment, there are only  limited labelled data of the target task available. In \cite{yosinski2014transferable}, it is reported that given a small dataset overfitting may happen if all parameters are re-trained, hence we only re-train partial parameters by freezing the remaining parameters to implement the fine-tuning. To be specific, we assume the number of neural network layers is $L$ and divide the pre-trained model into two parts. We set the first $L-1$ layers as the extractor, which is used to extract features of the problem and the last FC layer as the learner  which is used to refine the network in the target domain. We assume that the extractor part is non-trainable and only the learner part is trainable when the network is trained using the adaption dataset. Then, the parameter of FC layer of the pre-trained network can be updated by using either \eqref{tranparaupdate} or ADAM on the adaption dataset $\mathbb{D}_{\mathrm{Ad}}(\cdot)$. After finishing the training and fine-tuning steps, we obtain the adapted network model which can be used to predict the uplink power allocation coefficient on $\mathbb{D}_{\mathrm{Te}}(\cdot)$. The proposed DTL algorithm is summarized in Algorithm 1  which includes pre-training and fine-tuning stages.
\begin{table}\label{Algorithm2}
\hrule
\vspace{1mm} 
\noindent \textbf{Algorithm 1:}  The proposed offline adaptation algorithm based on DTL. \label{Table: Table II}
\vspace{1mm}
\hrule
\vspace{1mm}
$\hspace*{2mm}$\textbf{Input:} Learning rate $\alpha$ and $\beta$, batch size $N_b$, training dataset $\{\mathbb{D}_{\mathrm{Tr}}(k)\}_{k=1}^{N_{Tr}}$, adaptation dataset $\{\mathbb{D}_{\mathrm{Ad}}(k)\}_{k=1}^{N_{Ad}}$\\
$\hspace*{2mm}$\textbf{Output:} Learned network parameter $\mathbf{\theta}$
\vspace{1mm}
\hrule
\vspace{2mm}
$\hspace*{4mm}$$\mathbf{Pre-training}$
\begin{enumerate}
\item Randomly initialize the network parameter $\mathbf{\theta}$
\item Initialize the step: $t=0$
\item \textbf{while} not done \textbf{do}
\item $\hspace*{3mm}$ Randomly select $N_b$ sample pairs form $\{\mathbb{D}_{\mathrm{Tr}}(k)\}_{k=1}^{N_{Tr}}$ to compose a batch task
\item $\hspace*{3mm}$ $t\leftarrow t+1$
\item $\hspace*{3mm}$ Update the network parameter by $\mathbf{\theta}_t\leftarrow\mathbf{\theta}_{t-1}-\alpha\nabla_{\mathbf{\theta}_{t-1}}\mathrm{Loss_{\mathbb{D}_{Tr}}}
    (\mathbf{\theta}_{t-1})$ or by ADAM optimizer
\item \textbf{end while}
\end{enumerate}
\vspace{1mm}
\hrule
\vspace{2mm}
$\hspace*{4mm}$$\mathbf{Fine-tuning}$
\begin{enumerate}
\item Initialize $\tilde{\mathbf{\theta}}\leftarrow\mathbf{\theta}$
\item \textbf{for} $j=1,\ldots,G_{Ad}$ \textbf{do} 
\item $\hspace*{3mm}$ Update the parameter of the FC layer in $\tilde{\mathbf{\theta}}$ by using ADAM
\item \textbf{end}
\end{enumerate}
\hrule
\end{table}

\subsection{Meta Learning Algorithm}\label{metaalgorithmsect}
Different from transfer learning, meta learning aims to learn the best learning strategy, which is used to acquire an inductive bias for the entire class of tasks of interest for fast adaptation \cite{finn2017model}. We will design an offline meta learning algorithm to achieve faster and better adaptation than DTL in dynamic environments based on the MAML algorithm proposed in \cite{finn2017model}.
\subsubsection{Definition of Task}
Since the goal of the MAML algorithm is to train an efficient parameter initialization based on the multiple tasks, we define and form tasks before using MAML to design our algorithm. In our algorithm, a task is defined as a prediction process of uplink power allocation from channel realizations in a chosen dataset. Each such dataset is composed of training data and validation data for a particular task. We define a task set $\{\mathcal{T}_{mt}(k)\}_{k=1}^{K_{mt}}$, which includes $K_{mt}$ tasks. Each task in the task set is formed by randomly selecting training data and validation data from the meta training dataset $\mathbb{D}_{\mathrm{Tr}}(\cdot)$.
\subsubsection{Definition of Dataset}
Channel realizations and the associated optimal uplink power allocation vectors are involved in both training data and validation data of each task. The set of training data is defined as the support set $\mathbb{D}_{mts}(\cdot)$ and the set of validation data is defined as the query set $\mathbb{D}_{mtq}(\cdot)$. The support set and query set include $N_s$ and $N_q$ labelled data, respectively. We define the set used for adaption as the adaption dataset $\mathbb{D}_{Ap}(\cdot)$, which includes $N_{Ad}$ sample pairs. Note that the distribution of channel realizations in $\mathbb{D}_{Ap}(\cdot)$ is different from the distribution in $\mathbb{D}_{\mathrm{Tr}}(\cdot)$.
\subsubsection{Meta-training Stage}
\begin{figure}[ht]
\centering
\includegraphics[width=3.8in]{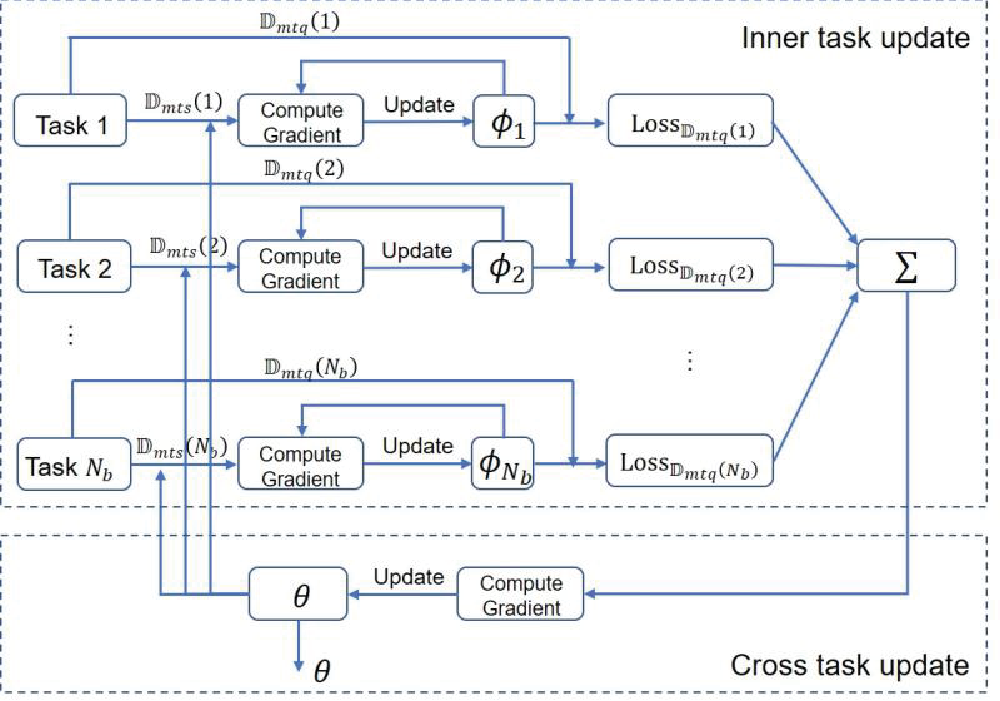}
\caption{The workflow of the meta learning.}
\label{meta_flow}
\end{figure}
The MAML algorithm uses two iterative processes, inner-task update and cross-task update, to generate the parameter initialization with the good generalization ability. Inner-task update is used to optimize the neural network parameter of each task, and cross-task update is used to optimize the global neural network based on the sum of the loss functions of all tasks. In order to efficiently solve the mismatch issue of beamforming design, these two iterative processes are adopted to design our meta learning based adaptive beamforming algorithm. In Fig. \ref{meta_flow}, the workflow of two iterative training processes is provided to explain the training process of our meta algorithm. In the following, we use a batch of tasks (include $N_b$ tasks) as an example to introduce the two processes in Fig. \ref{meta_flow}. The same neural network architecture is used in inner-task update and cross-task update.

Inner-task update is a process of training the neural network parameters of each task in the related batch. The goal of each task is to optimize its own neural network parameter on its support set via the global network parameter. The objective of each task is achieved by minimizing the loss function based on supervised learning. Although the objective function of each task is the same, the dataset used to achieve the goal of each task is different. The objective function of each task can be expressed as
\begin{align}\label{taskoptipro}
\phi_k=\arg\min_{\phi_k}\mathrm{Loss}_{\mathbb{D}_{mts}(k)}(\phi_k), ~~k=1,\ldots,N_b,
\end{align}where $\phi_k$ is the neural network parameter of task $k$ and it is initialized by the global network parameter $\mathbf{\theta}$, $\mathbb{D}_{mts}(k)$ is the support set of task $k$. Since the loss function $\mathrm{Loss}_{\mathbb{D}_{mts}(k)}$ in \eqref{taskoptipro} is represented by the MSE between the predicted value and the true value shown in \eqref{loss}, such loss function is differentiable. Hence, the gradient descent technique can be used to solve the optimization problem in \eqref{taskoptipro}. Multiple gradient updates are considered to update the parameter of each task rather than one gradient update originally proposed in \cite{finn2017model}. Notice that the neural network parameter of each task is independently updated. As we can see from Fig. \ref{meta_flow}, the updating process of the neural network parameter for each task on its support set is parallel. Since the objective function of each task is the same and the updating processes of the parameters for all tasks are parallel, we use task $k$ as an example to explain the updating process of its own neural network parameter on the related support set $\mathbb{D}_{mts}(k)$. The neural network parameter of all tasks is initialized by the global parameter $\mathbf{\theta}$. By using the gradient descent technique, the neural network parameter $\phi_k$ of task $k$ can be estimated by
\begin{align}\label{multipleupdat0}
\mathbf{\phi}_k^{(0)} = \mathbf{\theta}-\beta\nabla_{\mathbf{\theta}}\mathrm{Loss}_{\mathbb{D}_{mts}(k)}(\mathbf{\theta}),
\end{align}where $\beta$ is the learning rate of the  inner-task update, the superscript $0$ of $\mathbf{\phi}_k$ denotes the first iteration of gradient update. When the number of iterative steps is greater than one, the parameter $\mathbf{\phi}_k$ of task $k$ is updated by calculating the gradient of the loss function over its own parameter obtained at the previous iterative step, where is given by
\begin{align}\label{multipleupdati}
\mathbf{\phi}_k^{(i)} = \mathbf{\phi}_k^{(i-1)}-\beta\nabla_{\mathbf{\phi}_k^{(i-1)}}\mathrm{Loss}_{\mathbb{D}_{mts}(k)}(\mathbf{\phi}_k^{(i-1)}),
\end{align}where the superscript $i$ of $\mathbf{\phi}_k$ is the index of the iteration step and $i = 1, \ldots, G_{in}$, $G_{in}$ is the number of inner iterative steps. The `compute gradient' function in Fig. \ref{meta_flow} is used to compute the gradient of the loss function in \eqref{multipleupdat0} and \eqref{multipleupdati}. The repeated updating processes are represented by $\mathbf{\phi}_k,k=1,\ldots,N_b$, which is fed back to `compute gradient' in Fig. \ref{meta_flow}. The neural network parameter of task $k$ can be updated as $\mathbf{\phi}_k=\mathbf{\phi}_k^{G_{in}}$ when the number of iterations arrives at the final step $G_{in}$. Notice that the loss function $\mathrm{Loss}_{\mathbb{D}_{mts}(k)}(\mathbf{\phi}_k),\forall_k$ of each task is unknown and needs to be estimated on its support set $\mathbb{D}_{mts}(k)$ at each iterative step. When all tasks in the batch finish their iterations, the loss function can be considered as a metric to evaluate the trained parameter of each task on the related query set $\mathbb{D}_{mtq}(\cdot)$. These loss functions can be used to optimize the global network parameter $\mathbf{\theta}$ in cross-task update, which is described in the following part.

Cross-task update is a process of optimizing the global network parameter $\mathbf{\theta}$ based on the sum of the loss functions of all tasks in the batch. As mentioned in the inner-task update process, the loss functions of all tasks in the batch can be estimated based on the neural network parameter of the related tasks and their query sets when the maximum iteration step is achieved. Such loss functions can be added together to form the loss function used to optimize the global network parameter $\mathbf{\theta}$. This process is implemented by the sum function in Fig. \ref{meta_flow}. The objective function of optimizing the global network parameter $\mathbf{\theta}$ on a batch of tasks can be expressed as
\begin{align}\label{metaoptim}
\mathbf{\theta}=\arg\min_{\mathbf{\theta}}\sum_{k=1}^{N_b}\mathrm{Loss}_{\mathbb{D}_{mtq}(k)}(\phi_k),
\end{align}where $\mathbb{D}_{mtq}(k)$ is the query set of task $k$. Similar to inner-task update, the gradient descent technique can be used to update $\mathbf{\theta}$ in \eqref{metaoptim}, which is given by
{\begin{align}\label{metaupdate}
\mathbf{\theta}\leftarrow\mathbf{\theta}-\alpha\nabla_{\mathbf{\theta}}\sum_{k=1}^{N_b}\mathrm{Loss}_{\mathbb{D}_{mtq}(k)}(\mathbf{\phi}_k),
\end{align}}where $\alpha$ is the learning rate of cross-task update. Notice that there exists the chain rule when calculating the gradient of the loss function of each task in \eqref{metaupdate} since the neural network parameter of each task is updated at each iteration based on the updated parameter of this task at the previous iteration. Hence, the update of the neural network parameter in each iterative step needs to compute the gradient with respect to the parameter of the previous iterative step when computing the gradient of the loss function with respect to $\mathbf{\theta}$, which can be expressed as $\frac{\partial\mathrm{Loss}_{\mathbb{D}_{mtq}(k)}(\mathbf{\phi}_k)}{\partial(\mathbf{\phi}_k)}=\frac{\partial\mathrm{Loss}_{\mathbb{D}_{mtq}(k)}(\mathbf{\phi}_k^{G_{in}})}
{\partial(\mathbf{\phi}_k^{G_{in}})}\cdot\frac{\partial(\mathbf{\phi}_k^{G_{in}})}{\partial(\mathbf{\phi}_k^{G_{in}-1})}\cdot
\frac{\partial(\mathbf{\phi}_k^{G_{in}-1})}{\partial(\mathbf{\phi}_k^{G_{in}-2})}\cdot\ldots\cdot\frac{\partial(\mathbf{\phi}_k^0)}{\partial\mathbf{\theta}}$. It indicates that the MAML algorithm needs an additional backward pass since it involves a gradient through a gradient process. As shown in Fig. \ref{meta_flow}, the updated global network parameter $\mathbf{\theta}$ is considered as the initialized parameter of the tasks in the next batch and will be continuously updated. The algorithm will move to the next training step when the global network parameter in all batches completes the updating process by alternating inner-task update and across-task update in Fig. \ref{meta_flow}. An efficient parameter
initialization $\mathbf{\theta}$ will be obtained when the training is completed.

Different from the joint training method, which optimizes the neural network parameter based on the loss function of the single model shown in \eqref{jointupdate}, the proposed meta learning algorithm optimizes the model parameter via the loss functions of multiple tasks on their own model shown in \eqref{metaupdate}. Using multiple models will improve the generalization ability. Compared to the joint training method, the proposed MAML-based learning algorithm can generate the parameter initialization, which has better generalization ability and can help any task from the same distribution to achieve their optimal parameter more efficiently.

\subsubsection{Meta-adaption Stage}
Based on the initial global network parameter $\mathbf{\theta}$ obtained from the above meta-training stage, the network parameter will be updated using the adaptation dataset $\mathbb{D}_{Ap}(\cdot)$ to achieve fast adaptation to the new task. We set the number of adaptation steps as $G_{Ad}$. Before implementing the adaptation, we initialize the adaptation parameter $\phi_{Ap}$ as $\mathbf{\theta}$, which is obtained through the meta training stage. In the $j$th adaptation step, the network parameter $\phi_{Ap}$ can be updated as
\begin{align}\label{metaadapt}
\phi_{Ap}^{(j+1)}\leftarrow\phi_{Ap}^{(j)}-\beta\nabla_{\phi_{Ap}^{(j)}}\mathrm{Loss}_{\mathbb{D}_{Ap}}(\mathbf{\phi}_{Ap}^{(j)}).
\end{align}The iteration will finish when the stoping criterion is achieved. Full details for implementing meta-learning and meta-adaptation are summarized in Algorithm 2.

{\em \underline{Comparison of transfer learning and meta leaning}}: Transfer learning and meta learning both have the training and adaption stages. Although they have the same objective of achieving fast adaption, the strategies used in the training and adaption stages are different. Hence, transfer learning is not a special case of meta learning. Meta learning uses two iterative procedures to train the model, which means that it needs two backward passes in the training stage. However, transfer learning uses one backward pass to train the model in the training stage. In the adaption stage, meta learning re-trains all parameters on the new task whereas transfer learning only re-trains the parameter of the last layer while retaining the rest parameters.
\begin{table}\label{Algorithm2}
\hrule
\vspace{1mm} 
\noindent \textbf{Algorithm 2:}  The proposed offline adaptation algorithm based on meta-learning. \label{Table: Table II}
\vspace{1mm}
\hrule
\vspace{1mm}
$\hspace*{2mm}$\textbf{Input:}  {Learning rate $\alpha$ and $\beta$, batch size $N_b$, meta-training task set $\{\mathcal{T}_{mt}(k)\}_{k=1}^{K_{mt}}$, support set $\mathbb{D}_{mts}(k)_{k=1}^{K_{mt}}$ with $N_s$ labelled \\
$\hspace*{12mm}$data, query set $\mathbb{D}_{mtq}(k)_{k=1}^{K_{mt}}$ with $N_q$ labelled data, the number of inner-task update steps $G_{in}$, adaptation dataset $\mathbb{D}_{Ap}$, \\
$\hspace*{12mm}$and the number of adaptation  steps $G_{Ap}$}\\
$\hspace*{2mm}$\textbf{Output:} Learned initial network parameter $\mathbf{\theta}$
\vspace{1mm}
\hrule
\vspace{2mm}
$\hspace*{4mm}$$\mathbf{Meta-training}$
\vspace{-1mm}
\begin{enumerate}
\item Randomly initialize the network parameter $\mathbf{\theta}$
\item \textbf{while} not done \textbf{do}
\item $\hspace*{3mm}$ Randomly sample batch of task $\mathcal{T}_k$ from $\{\mathcal{T}_{mt}(k)\}_{k=1}^{K_{mt}}$
\item $\hspace*{3mm}$ \textbf{for} $\mathcal{T}_k, k=1,\ldots,N_b$ \textbf{do}
\item $\hspace*{6mm}$ Randomly sample support set $\mathbb{D}_{mts}(k)$ with $N_s$ sample pair and query set $\mathbb{D}_{mtq}(k)$ with $N_q$ sample pair from $\mathcal{T}_k$
\item $\hspace*{6mm}$ \textbf{for} $i=1,\ldots,G_{in}$ \textbf{do}
\item $\hspace*{9mm}$ Evaluate the gradient of the loss function of task $k$ on $\mathbb{D}_{mts}(k)$
\item $\hspace*{9mm}$ Update the task parameter based on \eqref{multipleupdat0} and \eqref{multipleupdati}
\item $\hspace*{6mm}$ \textbf{end}
\item $\hspace*{3mm}$ \textbf{end}
\item $\hspace*{3mm}$ Update the global network parameter $\mathbf{\theta}$ by \eqref{metaupdate} or by using ADAM optimizer
\item \textbf{end while}
\end{enumerate}
\hrule
\vspace{2mm}
$\hspace*{4mm}$$\mathbf{Meta-adaptation}$
\vspace{-1mm}
\begin{enumerate}
\item Initialize $\phi_{Ap}\leftarrow\mathbf{\theta}$
\item \textbf{for} $j=1,\ldots,G_{Ad}$ \textbf{do}
\item $\hspace*{3mm}$ Update all parameters in $\phi_{Ap}$ by using ADAM
\item \textbf{end}
\end{enumerate}
\hrule
\end{table}
\section{Online Meta-Learning Algorithm}\label{online_adaptation}
Although the proposed offline learning algorithms offer effective strategies to achieve fast adaptation to a new task, the design of the adaptation stage is based on the assumption that the dataset used for adaptation is available in advance and comes from a   stationary distribution. Under this assumption, the offline adaptation algorithms may not perform well in real-world wireless communication applications, such as vehicular communications in which the communications environment  may keep changing. This is because of two reasons: 1) the channel  is likely to become available sequentially since channel estimation methods normally need time to first obtain the channel statistics and then estimate the channel; 2) the channel may follow a non-stationary distribution as the environment continues to change. In order to enhance the adaptation capability of the proposed offline meta-learning algorithm in real-world applications, we propose an online adaptation algorithm in this section based on the online meta-learning framework introduced in \cite{finn2019online}.
\subsection{Online learning}
Online learning is a learning paradigm which uses the idea of continual learning on a non-stationary distribution of tasks over time \cite{hannan1957approximation}. The learner aims to sequentially learn the model parameter $\theta_t$ over all time slots. In order to measure the learning ability of a learner, the notion of regret is introduced, which is defined as difference between the cumulative loss of the learner $\sum_{t=1}^T\mathrm{Loss_{\mathbb{D}_t}}(\mathbf{\theta_t})$ and the cumulative loss of best single model $\sum_{t=1}^T\mathrm{Loss_{\mathbb{D}_t}}(\mathbf{\theta})$. The parameter $\theta_t$ is determined by the online learner, while and the parameter $\mathbf{\theta}$ is obtained by training the model based on the hindsight data \cite{shalev2012online}. The aim of online learning is to design an algorithm  which can make the corresponding regret grow  as slowly as possible.

FTL is a standard online learning algorithm \cite{hannan1957approximation}  which aims to update the parameter $\theta_t$ at slot $t$ based on the sum of the loss functions of the previous data $\mathcal{D}_{t-1}$. It can be expressed as
\begin{align}\label{FTLupda}
\theta_t=\mathrm{arg}\min_{\theta}\sum_{k=1}^{t-1}\mathrm{Loss_{\mathcal{D}_k}}(\mathbf{\theta}).
\end{align} 

\subsection{Online Meta-learning}
FTL may not learn an effective online model because it trains a single model on a single dataset from all prior time slots. In order to learn effective models to adapt to the non-stationary scenario, we consider to incorporate the meta-learning technique into FTL to design the online adaptation algorithm.
Note that we cannot directly apply the offline meta-training learning introduced in section III-B to design the online algorithm for the following reason.  In the offline scenario, the data used for adaptation are available in advance and come from a stationary distribution, which means all data in the adaptation set can be used to adapt the learning model. In the online scenario, however, the data used for adaptation  arrive  sequentially   and may come from  a non-stationary distribution, which means we cannot use the whole data in the adaptive set to implement the adaptation like the offline scenario. We need to use the cumulative data to implement adaptation  in an online manner. Based on the aforementioned difference, in the following we will provide   details for the design of our online meta adaptive algorithm.

Similar to the offline meta learning algorithm in section \ref{metaalgorithmsect}, the proposed online meta algorithm involves two processes of calculating the gradient in the meta training phase. The first gradient is used to update the task-specific parameter based on the network parameter. The second gradient is used to update the network parameter based on the updated task-specific parameters. To implement online learning, we assume that data received at each time slot is a task for  adaptation in subsequent time slots, and each task includes $N$ input/output pairs for each user. We use $\mathcal{T}_t$ to denote the task of the time slot $t$.  Then, we define an empty task set $\mathcal{B}_t$ to store the data of the task $\mathcal{T}_t$ at the time slot $t$. Notice that there is no training process at the beginning $t=0$. In the following, we use time slot $t>1$ as the example to describe the online learning process of the proposed algorithm.  At the beginning of the time slot $t$, the algorithm uses the task set $\mathcal{B}_t$ to store the sample pairs of $\mathcal{T}_t$ as the data of the task $\mathcal{T}_t$ arrives. The algorithm samples a minibatch of tasks with size $N_{task}$ from previous task sets $\{\mathcal{B}_{k},~k=0,\ldots,t-1\}$. For each task in the minibatch, we sample its training set $\mathcal{D}_k^{tr}$ with $N_{tr}$ sample pairs and validation set $\mathcal{D}_k^{val}$ with $N_{val}$ sample pairs from the corresponding task set $\mathcal{B}_k,~ k=0,\ldots,t-1$. {In the following, we will describe two iterative processes, one is the process to update the task-specific parameter and the other one is to update the network parameter at the time slot $t$.} First, we use task $\mathcal{T}_k$ as an example to describe the updating process of the task-specific parameter. Based on the training set $\mathcal{D}_k^{tr}$, the task-specific parameter $\phi_k$ for task $\mathcal{T}_k$ can be updated by the stochastic gradient descent method as follows:
\begin{align}\label{taskupda}
\phi_k\leftarrow\theta_t-\beta\nabla_{\theta_t}\mathrm{Loss_{\mathcal{D}_k^{tr}}}(\theta_t),
\end{align}where $\beta$ is the learning rate, $\mathrm{Loss}$ is the MSE loss function provided in \eqref{loss}. {$\theta_t$ is the network parameter at the time slot $t$, which is used to initialize the task-specific parameter $\phi_k$ of the task $k$ at the beginning of the updating process.} The equation in \eqref{taskupda} is used for the first gradient descent update on the task parameter $\phi_k$.  If multiple gradient descent updates are used, the updating equation after the first update is given by
\begin{align}\label{taskupdasec}
\phi_k^{(j)}\leftarrow\phi_k^{(j-1)}-\beta\nabla_{\phi_k^{(j-1)}}\mathrm{Loss_{\mathcal{D}_k^{tr}}}(\phi_k^{(j-1)}),
\end{align}where superscript $j$ of $\mathbf{\phi}_k$ is the index of the iterative step and $j = 2, \ldots, N_{in}$.  {Second, we move to estimate the network parameter once the updating process of the task-specific parameter for each task in the minibatch is finished.} Note that the same task may appear several times in a minibatch. Hence, we use index $Z_k\in[0,~N_{task}]$ to record the number of appearance times of task $\mathcal{T}_k$ in the corresponding minibatch. Based on the updated task-specific parameter $\phi_k^{N_{in}}$ of each task in the minibatch, the shared network parameter $\theta_t$ at the time slot $t$ can be updated using the validation set of the corresponding tasks as follows:
 {\begin{align}\label{networkpar}
\mathbf{\theta}_t\leftarrow\theta_t-\alpha\nabla_{\theta}\sum_{k=1}^{t-1}Z_k\mathrm{Loss_{\mathcal{D}_k^{val}}}(\phi_k^{N_{in}}),
\end{align}}where $\alpha$ is the learning rate. Once the iterative procedure of updating the network parameter $\theta_t$ of time slot $t$ is finished, we  adapt the trained model   using the current received the data in $\mathcal{B}_t$. The process is repeated in  the next time slot $t+1$. Full details of the online meta adaptation are summarized in Algorithm 3. We use the network parameter obtained from offline meta learning as the initialized network parameter for the learning algorithm.

\begin{table}\label{Algorithm3}
\hrule
\vspace{1mm}
\noindent \textbf{Algorithm 3:}  The proposed online meta-learning algorithm. \label{Table: Table III}
\vspace{1mm}
\hrule
\vspace{1mm}
$\hspace*{2mm}$\textbf{Input:} Learning rate $\alpha$ and $\beta$, offline trained network parameter $\mathbf{\theta}_{meta}$, empty task sets $\mathcal{B}_t$, $\forall_t$, the number of inner update \\
$\hspace*{12mm}$steps $N_{in}$, the number of adaptation steps $N_{ad}$, the number of minibatch size $N_{task}$, the received sample pairs $N$, the\\
$\hspace*{12mm}$sample pairs of training set and validation set $N_{tr}$ and $N_{val}$\\
$\hspace*{2mm}$\textbf{Output:} Learned network parameter $\mathbf{\theta}_t$ for each slot $t$
\vspace{1mm}
\hrule
\vspace{2mm}
\begin{enumerate}
\item \textbf{initialize} the network parameter $\mathbf{\theta}_1\leftarrow\mathbf{\theta}_{meta}$
\item \textbf{for} $t=0,\ldots$ \textbf{do}
\item $\hspace*{3mm}$ \textbf{if} $t=0$
\item $\hspace*{7mm}$ $\mathcal{B}_0\leftarrow\{\mathbf{h}_0, \mathbf{q}_0\}$ of task $\mathcal{T}_0$
\item $\hspace*{3mm}$ \textbf{else}
\item $\hspace*{7mm}$ $\mathcal{B}_t\leftarrow\{\mathbf{h}_t, \mathbf{q}_t\}$ of task $\mathcal{T}_t$
\item $\hspace*{7mm}$ \textbf{while} not done \textbf{do}
\item $\hspace*{11mm}$ random sample a minibatch of tasks and the corresponding set $\mathcal{D}_{k}^{tr}$ and $\mathcal{D}_{k}^{val}$ from $\mathcal{B}_k$, $k=0,\ldots,t-1$,
\item $\hspace*{12mm}$ \textbf{for all} sampled task \textbf{do}
\item $\hspace*{15mm}$ \textbf{for} $j=1,\ldots, N_{in}$ \textbf{do}
\item $\hspace*{19mm}$ update the parameter of each task by \eqref{taskupda} and \eqref{taskupdasec}
\item $\hspace*{15mm}$ \textbf{end for}
\item $\hspace*{11mm}$ \textbf{end for}
\item $\hspace*{11mm}$ update the shared network parameter $\mathbf{\theta}_t$ by using \eqref{networkpar}
\item $\hspace*{7mm}$ \textbf{end while}
\item $\hspace*{7mm}$ $\mathbf{--------------------Adaptation-------------------}$
\item $\hspace*{7mm}$ \textbf{initialize} $\theta_{ad}\leftarrow\theta_t$
\item $\hspace*{7mm}$ \textbf{for} $n=1,\ldots, N_{ad}$ \textbf{do}
\item $\hspace*{11mm}$ update $\theta_{ad}$ on current task set $\mathcal{B}_t$: $\theta_{ad}\leftarrow\theta_{ad}\beta\nabla_{\theta_{ad}}\mathrm{Loss_{\mathcal{B}_{t}}}(\theta_{ad})$
\item $\hspace*{7mm}$ \textbf{end for}
\item $\hspace*{7mm}$ $\theta_{t+1}\leftarrow\theta_t$
\item $\hspace*{3mm}$ \textbf{end if}
\item \textbf{end for}
\end{enumerate}
\hrule
\end{table}

\section{Simulation Results}\label{simu}
In this section, numerical simulations are carried out to evaluate the advantages of the proposed adaptive beamforming optimization algorithms  for different wireless communications scenarios. A MISO downlink system with one BS and multiple users is considered. The main simulation parameters are set as follows: carrier frequency is 2.9 GHz, bandwidth is 20 MHz, noise power spectral density is -174 dBm/Hz,  the learning rates are $\alpha=0.001$, $\beta=0.01$, the iterative steps are $G_{in}=G_{ad}=N_{in}=N_{ad}=20$, and the batch size is $N_b=20$. Other specific parameters including the number of antennas at the BS $M$, the number of users $K$ and the transmit power $P$ are provided in each figure. All input data and corresponding labels are generated by using MATLAB. Tensorflow 2.0.0 and Keras 2.3.1 are used to  implement the proposed DTL algorithm. PyTorch 1.4.0 is used to implement the proposed meta-learning algorithms. All simulation results are generated by using a computer with Intel i7-7700HQ CPU and 8 GB RAM. 

In our simulation, we assume that the BS can obtain perfect CSI based on channel estimation or feedback. Each sample pair in all datasets is composed of channel realization and the uplink power allocation vector. For each channel realization, we can generate its associated uplink power allocation vector by solving the uplink problem \eqref{uplproblem}. Channel realizations of the testing dataset and the adaption dataset come from the same distribution. The distribution of channel realizations in the testing dataset is different from that  of the training dataset. The channel realizations in training dataset used for transfer learning and meta-learning algorithms are generated by using three small-scale fading channel models: Rayleigh model with distribution $\mathcal{CN}(\mathbf{0}, \mathbf{I}_M)$, Rician model with rician factor 3, and Nakagami model with fading parameter 5 and the average power gain 2. We generate 5000 channel samples for each of the three small fading models. Hence, the training dataset includes 15000 sample pairs. For meta-learning algorithm, we randomly sample labelled data from 15000 sample pairs to construct 1500 tasks, $K_{mt}=1500$. Each task includes $N_s=50$ training sample pairs and $N_q=50$ validation sample pairs. In addition, 5000 testing sample pairs are generated for each testing channel model, which uses different channel distribution from the training data. The DL network shown in Fig. \ref{bnn} includes eleven layers, one input layer, two CL layers, two BN layers, three AC layers, one flatten layer, one FC layer, and one output layer. The input size of the input layer is $2\times MK$. For the two CL layers, each CL layer applies 8 kernels of size $3\times3$, one stride, and one padding. The input size of the first CL layer equals to the size of the input data. The input size of the second CL layer and the output size of both CL layers equal to $2\times MK\times8$. Besides, ReLU and Sigmoid functions are adopted at the first two activation layers and the last activation layer, respectively. Adam optimizer is adopted \cite{kingma2014adam}. The exponential decay rates for the first moment estimates and the second moment estimates are set to $0.9$ and $0.999$, respectively. The epsilon of Adam is set to $10^{-8}$.

We consider the following four typical fading scenarios for testing the adaptation capability of the proposed learning algorithms:
\begin{itemize}
\item Large-scale fading case: this model is a typically fading model used in communications systems.
\item WINNER II indoor case: the WINNER II indoor office scenario specified in \cite{bultitude20074}.
\item WINNER II outdoor case: a typical WINNER II  urban scenario specified in \cite{bultitude20074}.
\item Vehicular case: we adopt an urban vehicle-to-infrastructure (V2I) scenario defined in Annex A of 3GPP TR 36.885  \cite{3gpp885}.
\end{itemize}

For comparison, we introduce other three benchmarks, namely, the optimal solution, the BNN solution, and the joint learning solution. The definitions of all solutions for comparison are listed below:
\begin{itemize}
\item The optimal solution: this solution shows the optimal result of the problem in \eqref{SINRproblem} obtained by using the iterative algorithm proposed in \cite{schubert2004solution}. It serves as a performance upper bound for all other schemes.
\item The BNN solution: this solution \cite{xia2019deep} shows the predicted result, which is obtained based on the assumption that wireless channels in the training dataset follow the  same fading distribution as those in the testing dataset. It provides a performance upper bound for our proposed adaptive algorithms.
\item The transfer learning solution: this solution shows the adaptation result of  the proposed DTL algorithm in Section III-A.
\item The meta-learning solution: this solution shows the fast adaptation result of the proposed  meta-learning algorithm in Section III-B.
\item The joint training solution: this solution shows the result obtained by training the neural network using all available data without adaptation.
\end{itemize}
For fair comparison, we set the same convergence criteria for the iterative procedures of all schemes. In addition, we set the same size of training set of fine-tuning for transfer learning and meta-learning algorithms. The results and analysis are provided for each case below.
\subsection{Large-scale fading case}
\begin{figure}[ht]
\centering
\includegraphics[width=3.8in]{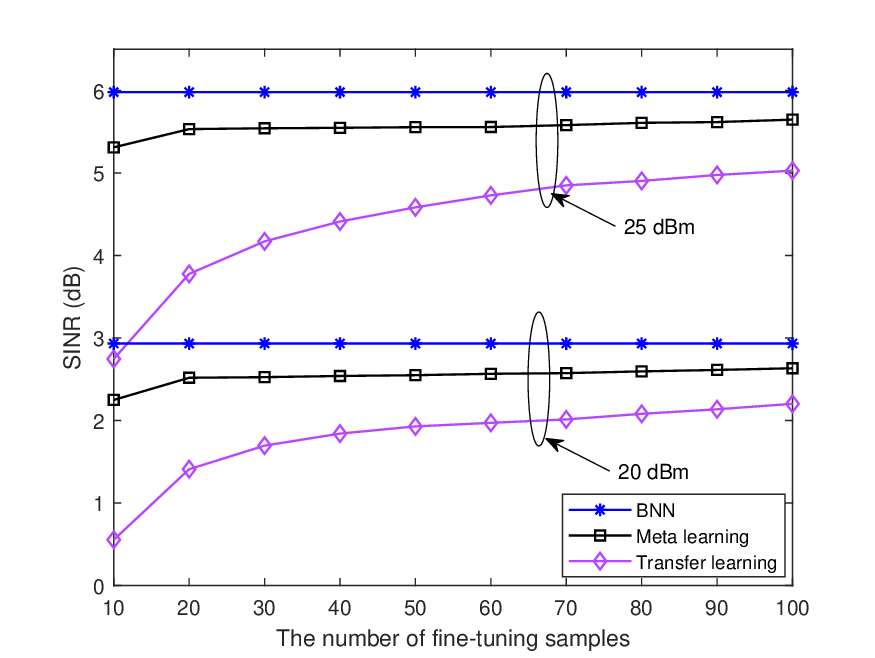}
\caption{The comparison of fine-tuning samples for meta and transfer learning when $M = 8, K = 8$.}
\label{fig1}
\end{figure}
In this case, the pathloss model is given by $\mathrm{PL}=128.1+37.6\log_{10}(d ~[km])$, where $d$ is the distance between a user and BS. The shadow fading follows the log-normal distribution with zero mean and $8$ dB standard deviation. The Rayleigh fading channel with zero mean and unit variance is adopted as the small-scale fading for this case. All users are randomly distributed within a disc with a radius of $500$ m. In order to choose the proper size of fine-tuning samples for the proposed two adaptation algorithms, we first investigate the effect of the number of fine-tuning samples used in the transfer learning and the meta-learning algorithms in Fig. \ref{fig1}. As the figure shows, the SINR increases when the number of fine-tuning samples increases for both algorithms under different transmit power settings. The SINR   generated by the proposed meta-learning algorithm almost converges by using only 20 fine-tuning samples. However, there is still an obviously gap (almost 1 dB for 25 dBm and 0.5 dB for 20 dBm of transmit power) between the meta-learning and transfer learning algorithms when the number of fine-tuning samples increases to 100. By considering the adaptation overhead and the SINR performance, we choose 20 samples for fine-tuning of the transfer-learning  and  the meta-learning algorithms in all testing channel models.

\begin{figure}
\centering
\subfigure[] 
{
	\begin{minipage}{3.1in}
    \label{fig2:a}
    \centering
	\includegraphics[width=3.1in]{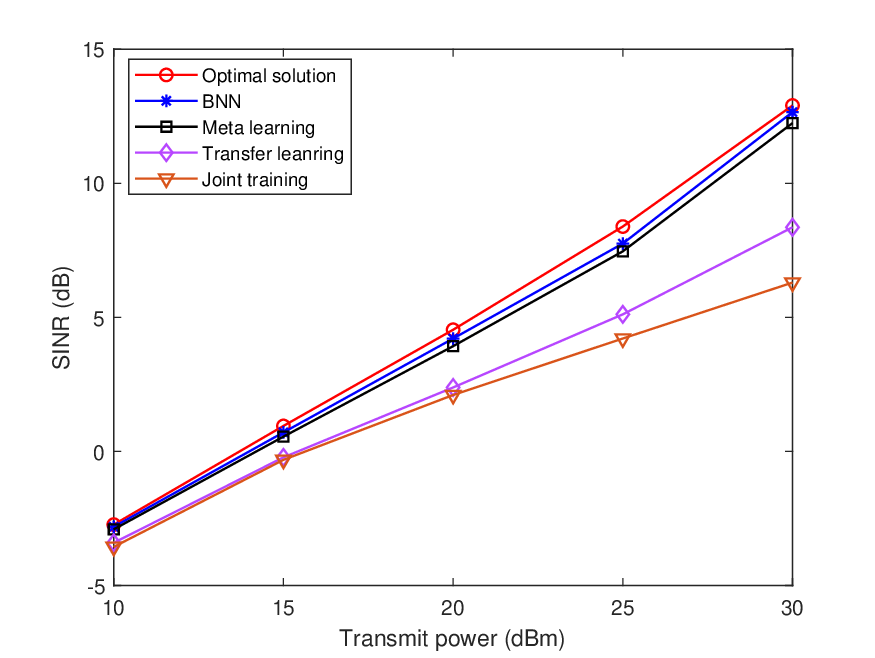}   

	\end{minipage}
}
\subfigure[]  
{
	\begin{minipage}{3.1in}
    \label{fig2:b}
    \centering
	\includegraphics[width=3.1in]{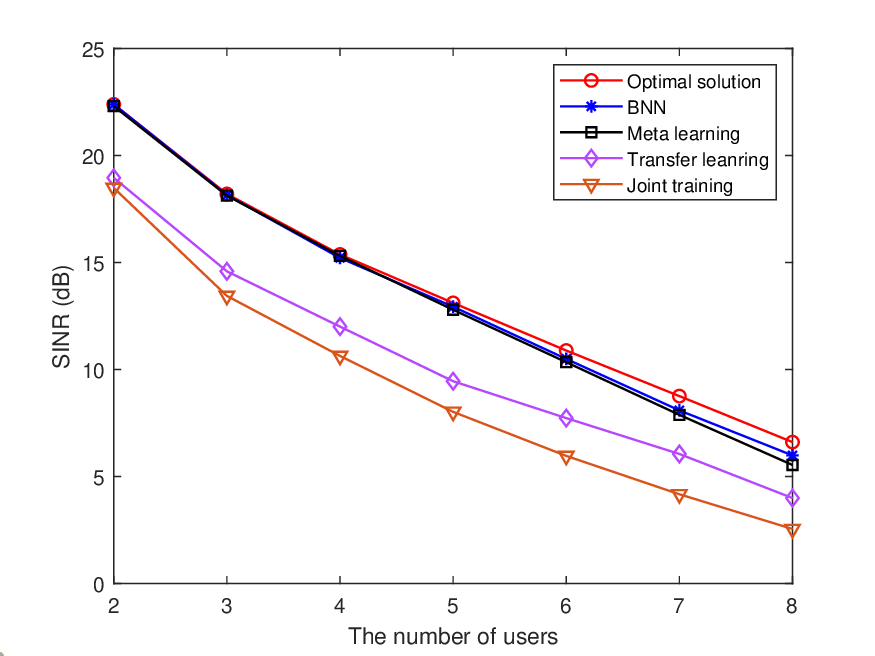}   
	\end{minipage}
}
\centering
\caption{The SINR performance comparison on large-scale case for different metrics: (a) transmit power when $M = 4$, $K = 4$ and (b) the number of users when $M = 8$, $P = 25$ dBm.}  
\label{fig2}   
\end{figure}
Based on the 20 fine-tuning samples, we demonstrate the adaptation capability of the proposed algorithms via the SINR performance using two different metrics in Fig. \ref{fig2}.  Fig. \ref{fig2:a} shows the effects of the transmit power on the SINR performance. As can be seen, the SINR increases as the transmit power increases for all schemes. The SINR result generated by the proposed meta-learning algorithm is very close to that of the BNN scheme which validates its effectiveness. It is observed that the performance gap between the transfer learning and the meta-learning is significantly enlarged as the transmit power increases. The joint training scheme without adaptation achieves the worst SINR compared to other schemes. In Fig. \ref{fig2:b}, the SINR performance becomes worse as the number of users increases. The proposed meta-learning algorithm still produces better result  which is close to the BNN scheme, compared to the transfer learning and the joint training scheme. It is interesting that there is an obviously reduction of the SINR performance gap between the meta-learning algorithm and the transfer learning algorithm when the number of users is greater than 6. This is because more channel features can be extracted and transferred by using the transfer learning algorithm as more users are involved. Overall, the results plotted in Fig. \ref{fig2} demonstrates that the proposed meta-learning algorithm provides an efficient beamforming adaptation solution.
\subsection{WINNER II indoor case}
\begin{figure}
\centering
\subfigure[]  
{
	\begin{minipage}{3.1in}
    \label{fig3:a}
    \centering
	\includegraphics[width=3.1in]{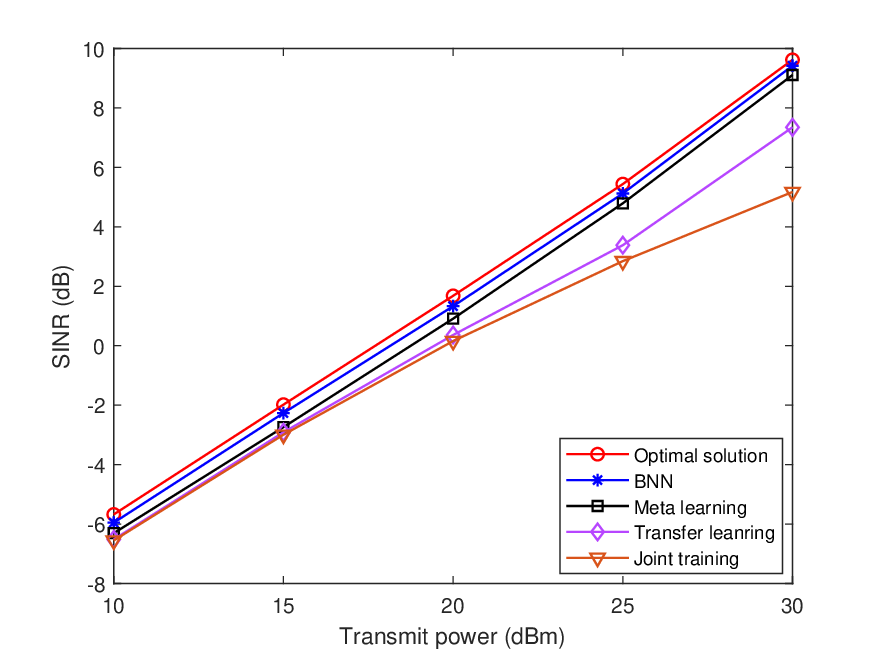}   

	\end{minipage}
}
\subfigure[]  
{
	\begin{minipage}{3.1in}
    \label{fig3:b}
    \centering
	\includegraphics[width=3.1in]{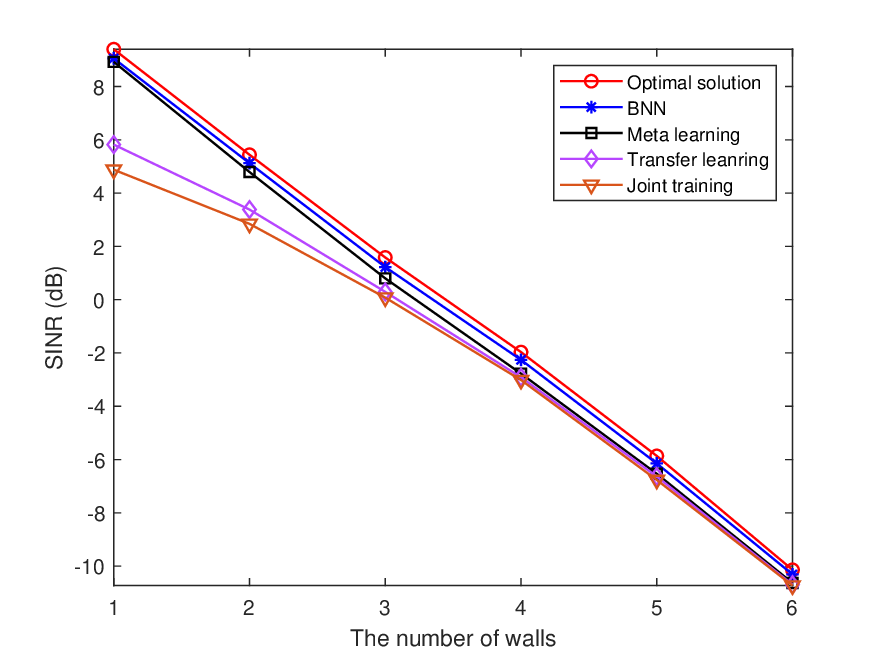}   
	\end{minipage}
}
\centering
\caption{The SINR performance comparison in the WINNER II indoor case when $M = 4$, $K = 4$: (a) transmit power when $n_w = 2$ and (b) the number of walls when $P = 25$ dBm.}  
\label{fig3}   
\end{figure}
In the WINNER II indoor case, we assume that the access point (AP) and users are located on the same floor. Users are randomly located between 10 m to 100 m away from the AP. We adopt the corridor-to-room scenario, in which only non line-of-sight (NLOS) path is considered due to the blocked light-of-sight (LOS) path. The pathloss is given by $\mathrm{PL}=43.8+36.8\log_{10}(d~[m])+20\log_{10}(f_c/5)+5(n_w-1)$, where $f_c$ is carrier frequency and $n_w$ is the number of walls. The standard deviation of shadow fading is $4$ dB. The adaptation capability of the proposed learning algorithms in the WINNER indoor case is investigated in Fig. \ref{fig3} based on two   factors: the transmit power at the AP  and the number of walls. Fig. \ref{fig3:a} shows that the adaptation capability of the proposed meta-learning algorithm in the indoor fading case is similar to that demonstrated in the large-scale case. Different from the results shown in Fig. \ref{fig2:a}, the SINR performance of the transfer learning algorithm is close to that of the meta-learning algorithm when the transmit power is smaller than 15 dBm in Fig. \ref{fig3:a}. This fact may indicate that the system needs to spend more power on overcoming the fading caused by the walls. Hence, the effects of the number of walls on the adaptation capability for the proposed algorithms are provided in Fig. \ref{fig3:b}. As can be seen from Fig. \ref{fig3:b}, the SINR performance decreases for all schemes when the number of walls between the user and AP increases. In addition, the performance gap between the meta learning and the transfer learning algorithm rapidly reduces as the number of walls increases.  
\subsection{WINNER II outdoor case}
\begin{figure}
\centering
\subfigure[]  
{
	\begin{minipage}{3.1in}
    \label{fig4:a}
    \centering
	\includegraphics[width=3.1in]{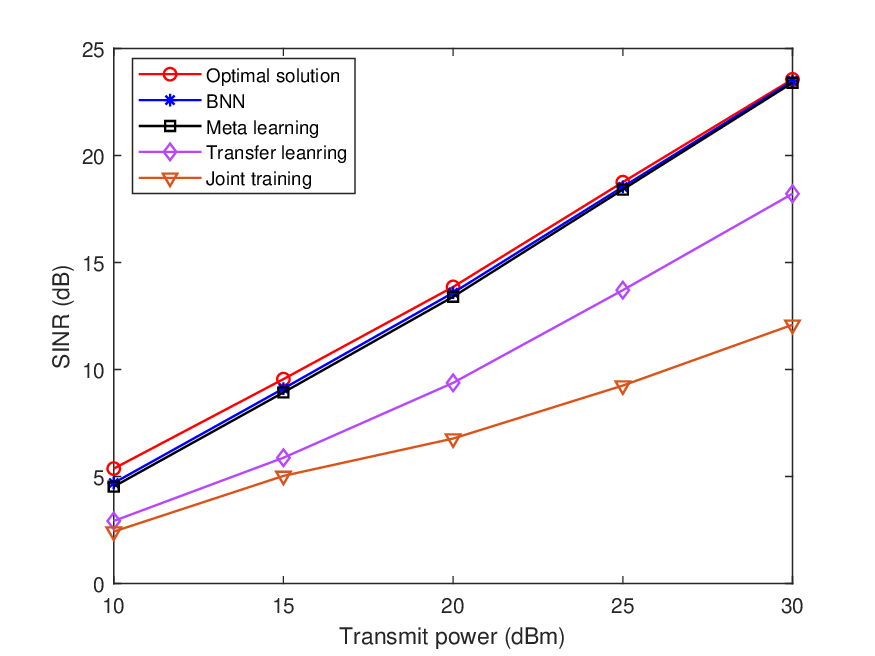}    
	\end{minipage}
}
\subfigure[]  
{
	\begin{minipage}{3.1in}
    \label{fig4:b}
    \centering
	\includegraphics[width=3.1in]{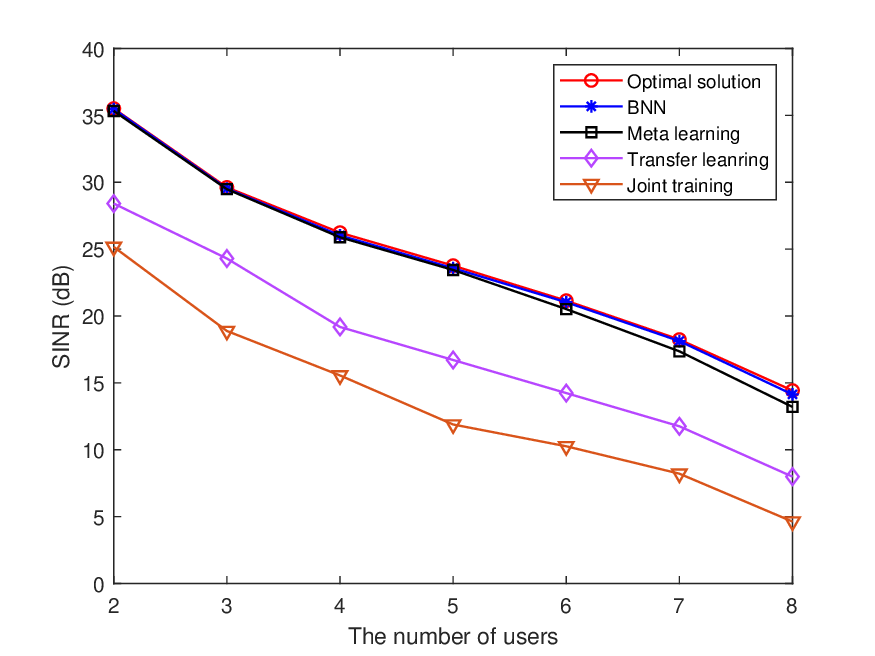}    
	\end{minipage}
}
\centering
\caption{The SINR performance comparison in the WINNER II outdoor case: (a) transmit power when $M = 4$, $K = 4$ and (b) the number of users when $M = 8$, $P = 25$ dBm.}  
\label{fig4} 
\end{figure}
In the WINNER II outdoor case, we assume that the BS is located in the cell center and covers a disc with a radius of $1000$ m. Users are randomly distributed between $100$ m to $1000$ m away from the BS. The pathloss and shadowing of LOS in WINNER B1 are adopted to generate the large-scale fading for this case. Fig. \ref{fig4} demonstrates the adaptation capability of the proposed learning algorithms in the WINNER II outdoor case through the SINR performance. As can be seen from Fig. \ref{fig4}, the proposed meta-learning algorithm achieves the highest SINR performance compared to the proposed transfer learning algorithm and the joint training algorithm as the transmit power and the number of users change. Compare with the SINR performance in Fig. \ref{fig2:a} and Fig. \ref{fig3:a}, the performance gap between the proposed transfer learning algorithm and the joint training algorithm is significantly increased when the transmit power is greater than 15 dBm in Fig. \ref{fig4:a}.
\subsection{Vehicular case}
\begin{figure}[ht]
\centering
\includegraphics[width=3in]{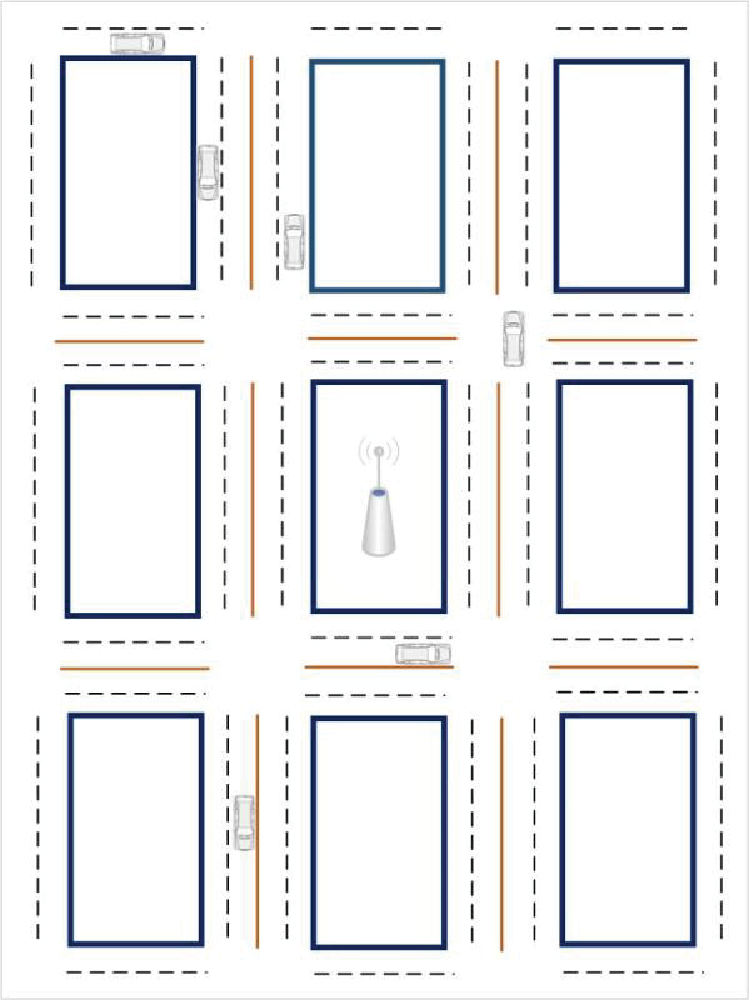}
\caption{Manhattan road grid $750~m\times 1299~m$ \cite{3gpp885}.}
\label{fig5}
\end{figure}
In the vehicular case, we use the Manhattan grid layout with the region size of $750 ~m\times 1299 ~m$ to set up a realistic V2I communication scenario as shown in Fig. \ref{fig5}. The size of each grid is $250 ~m\times 433 ~m$. There are two lanes in each direction for vehicles with 3.5 m lane width. The BS is located in the center of the layout. The vehicles are uniformly placed on each direction of the road. The probability of each vehicle to change its direction at the intersection is set to be 0.4. Each vehicle will change its direction when it arrives at the edge of the layout. We assume that the velocity of each vehicle is 60 km/h. The pathloss and shadowing adopted in this case are the same to those in the large-scale case. Besides, the antenna gains of the BS and each vehicle user are set to be 8 dBi and 3 dBi, respectively. The decorrelation distance is set to 50 m. We use Clarke's model introduced in \cite{tse2005fundamentals} to generate the small-scale fading of the moving vehicles.

\begin{figure}
\centering
\subfigure[]  
{
	\begin{minipage}{3.3in}
    \label{fig6:a}
    \centering
	\includegraphics[width=3.3in]{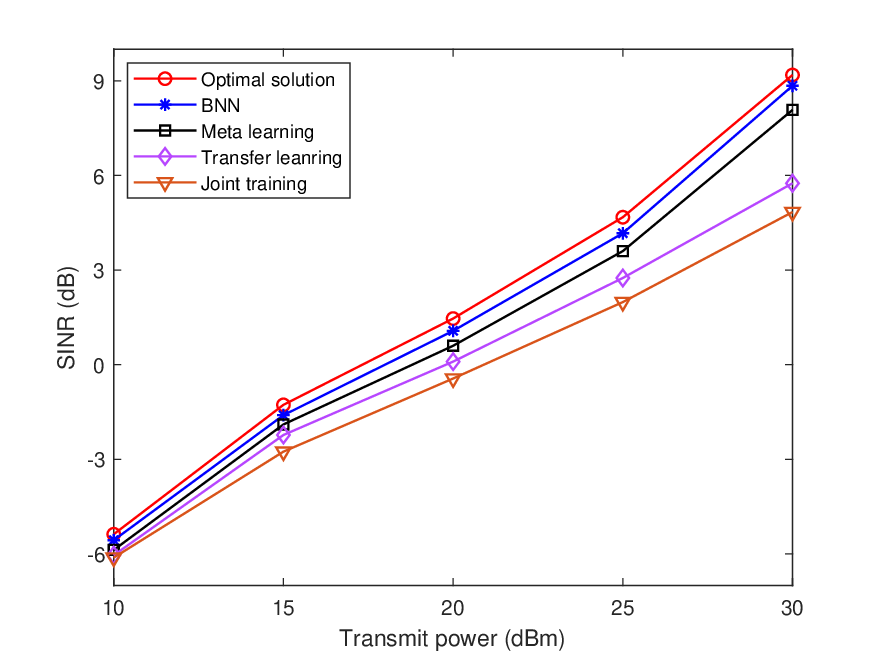}    
	\end{minipage}
}
\hspace{0.5in}	
\subfigure[]  
{
	\begin{minipage}{3.3in}
    \label{fig6:b}
    \centering
	\includegraphics[width=3.3in]{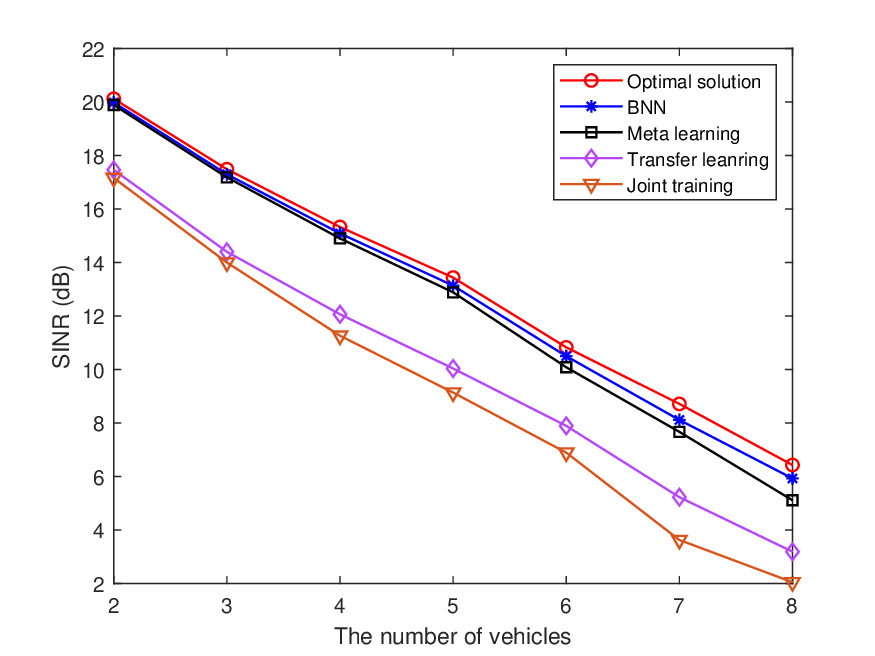}   
	\end{minipage}
}
\hspace{0.5in}
\subfigure[]
{
	\begin{minipage}{3.3in}
    \label{fig6:c}
    \centering
	\includegraphics[width=3.3in]{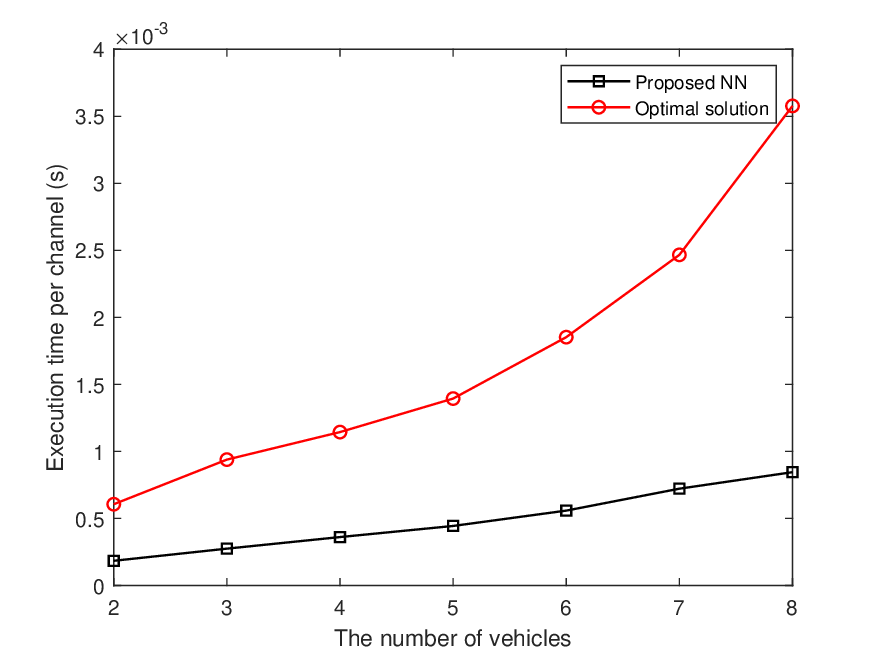}    
	\end{minipage}
}
\centering
\caption{The performance and complexity comparison of the proposed algorithms in the vehicular case: (a) SINR versus transmit power when $M = 4$, $K = 4$, (b) SINR versus the number of vehicles when $M = 8$, $P = 25$ dBm and (c) execution time per channel versus the number of vehicles when $M = 8$, $P = 25$ dBm.} 
\label{fig6}  
\end{figure}
The effects of the transmit power and the number of vehicles on the SINR performance are presented in Fig. \ref{fig6:a} and Fig. \ref{fig6:b}, respectively. For both factors, the SINR performance of the proposed meta-leaning algorithm is close to that of the BNN scheme. Fig. \ref{fig6:a} shows that the proposed meta-learning and transfer learning algorithms significantly outperform the joint training scheme when the transmit power is greater than 14 dBm. There exists an obvious  performance gap between the two learning algorithms when the transmit power is greater than 25 dBm. Fig. \ref{fig6:b} shows that the meta-learning algorithm significantly outperforms the transfer learning algorithm and the joint training method. The SINR performance gap between the transfer learning algorithm and joint training method becomes large as the number of vehicles increases. Similar to the above three fading cases, the proposed meta-learning algorithm   provides superior performance thanks to its fast adaptation   even in the moving scenario. Fig. \ref{fig6:c} shows the comparison of the execution time  between the proposed algorithms (same for the transfer learning and the meta-learning algorithms) and the optimal solution.   From the figure, we can see that the proposed learning algorithms requires much less time compared to the optimal solution. This is because no iterative process is used in the proposed learning algorithms to predict the beamforming solution.

In addition, we compare the fine-tuning execution time of the meta-learning algorithm and transfer learning algorithm in Table \ref{fine_tuning_time}. The results are obtained by using 20 fine-tuning samples and 20 iterative steps. In order to make the fair comparison between the proposed two algorithms, the adaption of the transfer learning algorithm is also implemented by using PyTorch. From  Table \ref{fine_tuning_time}, we can see that the execution time of the transfer learning algorithm to achieve adaption is less than that of the meta learning algorithm and this is because only part of the neural network needs to be re-trained in transfer learning.
\begin{table*}[ht]
  \centering
  \caption{Comparison of fine-tuning execution time between meta learning and transfer learning.}\label{fine_tuning_time}
  \begin{tabular}{|c|c|c|c|c|c|c|c|}
     \hline
     No. Users & 2&3&4&5&6&7&8\\
     \hline
     Meta-learning (ms)&95&97&100&102&106&108&113 \\
     \hline
     Transfer learning (ms)&74&78&80&85&87&90&95\\
     \hline
   \end{tabular}
\end{table*}
\subsection{Online learning}
\begin{figure}[ht]
\centering
\includegraphics[width=3.8in]{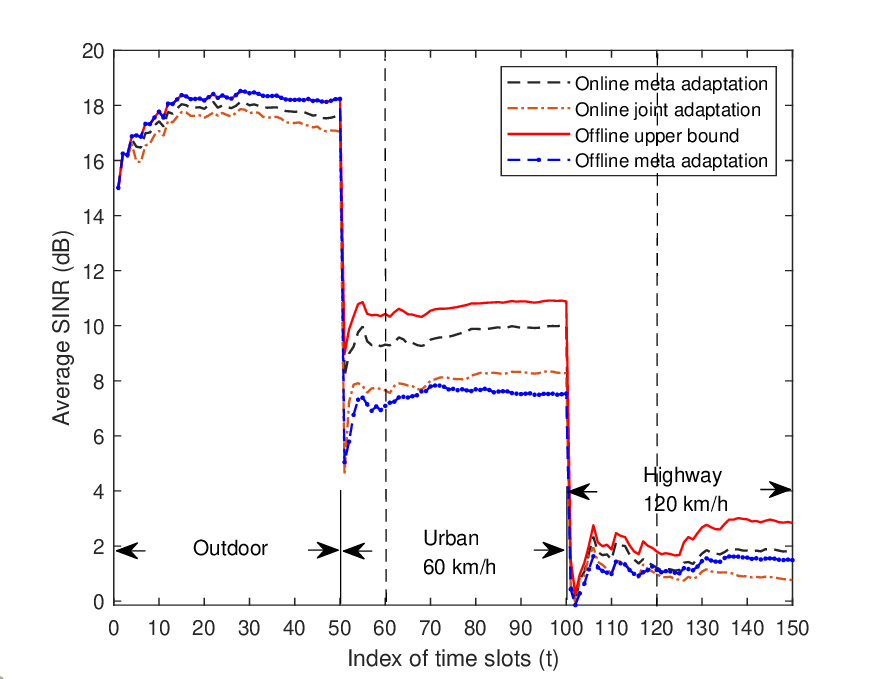}
\caption{Performance comparison between online and offline adaptation algorithm {$M = 4$, $K = 4$, $P = 25$ dBm}}
\label{fig7}
\end{figure}
In this case, we evaluate the performance of the proposed  online meta-learning algorithm in real-world non-stationary   scenarios. We consider mobile users travelling from outdoor to urban and then highway environments in the simulation to investigate the adaptation capability of the proposed online algorithm in the changing environments. The WINNER II outdoor and vehicular case are used to generate channel data for the outdoor and urban scenarios, respectively. The freeway case introduced in 3GPP TR 36.885 \cite{3gpp885} is used to generate channel data for the highway scenario. The number of lanes in each direction and the velocity of each vehicle are set as $3$ and $120$ km/h, respectively. The antenna gains of the BS and vehicle in the urban and highway scenario are set as 8 dBi and 3 dBi, respectively.   We set the minibatch $N_{task}=20$, the same sample pairs for training set and validation set $N_{tr}=N_{val}=4$. The results of the online joint adaptation algorithm is obtained by using the FTL method.   Fig. \ref{fig7} shows the adaptation performance comparison between the proposed offline and online learning algorithm over the whole communications period as the users move across different environments. Each simulation point in the figure is obtained by averaging all of the actual experimental points at the individual time slots over the previous time slots in the corresponding communications scenario. To implement simulation, we assume that five adaptation channels $N=5$ and ten testing channels of each user are received at each time slot. Each communications scenario lasts 50 time slots. For the offline meta adaptation case, the neural network model will be updated periodically at every 60 time slots based on the collected channel information during that period. For the offline upper bound case, we assume that the system knows the exact   environment   for all communications scenarios in advance and also knows when to update the model, so it provides a performance upper bound. The online joint adaptation uses all data until slots $t-1$ to train the model and then use the data at slot $t$ to fine tune the model. As we can see from the figure, there is an obvious  drop  on the average SINR for all cases when the communications scenario changes, which indicates that the changing communications scenario can significantly affect the system performance. There exists an obvious gap between the online algorithm and the offline upper bound, and the performance gap is obviously enlarged when the communication scenario changes from outdoor to urban, whereas the performance gap is slightly increased when the scenario changes from urban to highway. The reason is that urban and highway are both scenarios with high mobility and they share more similar channel statistics features. It is interesting to point out that the offline meta adaptation algorithm performs worse than the online joint method from the beginning of the urban scenario to the beginning of the highway scenario. The reason is that the offline meta adaptation algorithm still uses the trained model based on the previous scenario (outdoor/urban) to the new scenarios (urban/highway). The  difference in mobility between different environments causes the task mismatch issue for the offline meta learning algorithm before its periodic update. Whereas the offline meta adaptation algorithm outperforms the online joint method after updating its model in the highway case. This fact indicates that the offline algorithm heavily relies on the stationary environment. The proposed online meta adaptation algorithm has the best performance compared to the online joint method and the offline meta adaptation, and it can fast adapt to the new communications scenario by effectively making use of   sequential   data.

\section{Conclusion}\label{conc}
In this paper, we have proposed two offline learning algorithms to achieve fast adaptation on the beamforming design when the distribution of testing wireless environments changes. For the DTL algorithm, it utilizes the pre-trained model to re-train part of the parameter in order to achieve the adaption on the new environment. Different from the DTL algorithm, the meta learning algorithm aims to learn the parameter initialization, which can be used to achieve fast adaption to the new environment by re-training the whole neural network. In order to enhance the adaption capability of the proposed offline meta-learning algorithm in the real-world  scenarios, an online adaptive learning algorithm was proposed based on the meta-learning algorithm and the FTL approach. Simulation results demonstrated that both offline algorithms can achieve fast adaption when testing data come from a different distribution, and the proposed meta learning algorithm provides   better generalization ability by using slightly more adaptation time compared to the DTL algorithm. The online adaptive algorithm can significantly enhance the adaption capability of the proposed offline meta-learning algorithm in non-stationary scenarios. In this paper we only consider a single task distribution which is to optimize beamforming. In the future, we plan to use the lifelong learning technique \cite{thrun1998lifelong} to solve more complex problems in self-organizing networks, in which tasks may come from different distributions, so bidirectional learning, knowledge retention and accumulation are necessary.

\end{document}